\documentstyle[twoside,fleqn,espcrc2]{article}

\title{Multiparticle Dynamics 1997: Concluding Talk}

\author{J. D. Bjorken\thanks{Work supported by the Department of Energy under
contract number DE--AC03--76NSF00515.} \\[2ex]
Stanford Linear Accelerator Center, Stanford University,
Stanford, California 94309}
\begin{document}

\begin{abstract}
This contribution to the XXVII Symposium on Multiparticle Dynamics held in
Frascati, Italy, September, 1997 consists of the following subject matter: 
\begin{enumerate}
\item
Introductory generalities.
\item
 Brief mention of some of the contributions to the meeting.
\item
 More extended discussion of a few specialized topics.
\item
 Discussion of the FELIX initiative for a QCD detector at the LHC.
\end{enumerate}
\end{abstract}
\maketitle

\section{GENERALITIES}

In this meeting there were a variety of important and interesting
contributions, many if not most of which represent extrapolations of what
has transpired in previous meetings in this series. I have discovered in
fact that it is also true for this talk\cite{bj1}. But anyway I felt a
certain lack of focus in the course of the week, and it seems appropriate
here to first review the most basic question to be asked: why do we do
what we do in this field? In earlier times, when this series of meetings
was initiated, it was to find out something about the nature of the strong
force via multiparticle dynamics. But now at the fundamental level almost
everyone agrees that this is a solved problem: the strong interactions are
described by the Lagrangian and equations of motion of QCD. Only a few
voices cry out from the most distant and obscure wildernesses on the
planet, e.g. Milano, challenging the conventional wisdom. 

This means for me that it is of less and less significance to ``test QCD",
unless the test---usually a precise determination of $\alpha_s$---is
demonstrably superior to what has already been done. It becomes more and
more important to ``understand QCD", which means to uncover all its
consequences---especially those which lie beyond the methodology of
perturbation theory---and provide appropriate descriptions of these
phenomena. By ``appropriate" means that they are at least in principle
consistent with the general consequences of the underlying equations of
motion, and at best derivable from them. It also is of course the case,
almost by definition, that the best descriptions will account for the most
data, will provide reliable predictions, and will be concise and
comprehensive.

So the goal has not changed so much; in the first paragraph above, only
the phrase ``...the nature of the strong force..." has to be changed to
``...the nature of QCD...". 

In this respect, we may compare QCD with its namesake QED, especially in
the limit of HNET (heavy nucleon effective theory), i.e. atomic physics. 
Certainly the precision tests of QED are important, and remain so, just as
for QCD. But beyond perturbative QED lies the physics of the ozone layer
(atomic collisions and electromagnetic processes at very low energy),
metallurgy, superconductivity, organic chemistry, and the life sciences
themselves, just as examples. It is not necessary to approach these
subfields direct from the QED Lagrangian, although it is a splendid thing
to see how the derivations go. It is quite enough to savor them on their
own terms. 

It is hard to imagine that there ever will be the incredible richness of
the HNET/QED world reflected in the world of QCD. But there are plenty of
frontiers out there, with in my opinion the most interesting and exciting
of them those which are the most nonperturbative, and least under direct
theoretical control. To investigate them well demands, just as in
HNET/QED, a very close interplay between theory and experiment.

Among the most basic frontiers of QCD is the question of what happens at
much higher energy scales. The expectation is that there will be big
changes in the phenomenology of generic hadron-hadron collisions, even at
the qualitative level. This realization has been sharpened, at least for
me, very recently.  This is due to the emergence of a new initiative, FELIX, 
a detector proposed to comprehensively examine QCD phenomena at the LHC. 
I will return at some length to this in Section 4, because it is on my mind very 
much these days, and because it is an opportunity of great relevance to this field of
multiparticle dynamics. 

But there are many other QCD frontiers.  There is an important
large-distance frontier, something that can be called LQET (light-quark
effective theory). Here the chiral effective action is the appropriate
descriptive tool. There is a school of thought that it can be even
derived, more or less, from the basic short-distance QCD action via
integrating out the effects of instanton contributions.  Instantons arguably are
the driving mechanism of spontaneous breaking of the approximate chiral
symmetry of QCD\cite{ss}. A more open question is how in detail confinement of
color occurs; this is no doubt still the most important open frontier of QCD.

The large-distance theoretical frontier has an experimental counterpart. The
extraordinary technological advances made in the large collider-detector
community, both in high rate data acquisition and in precision tracking in
complex background environments, has not been applied in full force to low
energies. A fully modern detector for light-quark spectroscopy would play
an important role;  it is a pity that as far as I can see there is no such
device even being proposed. 

In between the short-distance and large-distance worlds of QCD lies
the boundary region, and this is an area of crucial importance to the
subject of multiparticle dynamics. Defining the energy and/or
distance scale of this boundary is vital, as well as its ``thickness":  how
much higher in energy scale or shorter in distance scale must one go
before perturbative methods become reliable? There is an increasing
tendency to push the boundary from the high energy perturbative side down
to very low energy/momentum scales, especially in multiparticle dynamics,
and I at least worry about how meaningful this is\cite{bj1}. I will return to this
question in Section 3. 

Other important frontiers include the small-$x$, BFKL frontier in
short-distance QCD, the problem of leading-particle behavior in
hadron-hadron collisions, almost everything concerning inelastic
diffraction, and even the parton structure of the nucleon.  While there is
a vast amount of information on the longitudinal distribution of partons
in an infinite-momentum nucleon, there is almost nothing on how the
different partons of a given configuration are correlated with each other. 
And there is very little information on how the partons are distributed
transversely, even at the inclusive level. For example, we tell people
that the proton is made of three (constituent)  quarks. Then we also tell
them (or should tell them) that there are an infinite number of
quark-partons in the infinite-momentum proton. Where are they in the
impact plane, relative to the location of the constituent quarks? Are they
mostly inside them or outside? I think the honest answer is that we do not
know the answer to such a simple question.  I favor the latter option. But
it remains the case that there is a frontier out there: what is the right
answer and how do we find it out? 

\section{SOME INTERESTING NEWS}
This is not a summary talk, and this section represents as close as it
will get to being one, being a sequence of advertisements of some of the
contributions that especially attracted my attention during the meeting.
Only a few will be elaborated upon here, and in any case the best way 
to find out what any of them is really about, is to go directly to them
and not to linger here expecting much more. It is also important to not
trust this list and go straight to the rest of the proceedings.

{\em NA50} (Giubellino\cite{**}): The evidence for an unusual mechanism for
$\psi$ suppression in heavy ion collisions is to my eye quite impressive.
What other properties of the collisions can be correlated with this abrupt
change with increasing $A$?  As always, I would like to understand the
space-time geometry better; it seems that the ion-ion final-state system
is already undergoing mainly spherical expansion at the $\psi$ formation
time for light-ion or noncentral collisions, while it may still be in a
more dominantly longitudinal expansion stage for $Pb-Pb$ collisions.  Can
this change in the geometry have an effect on the change in the amount of
suppression?

{\em Screwy gluons}: Bo Andersson's talk\cite{**} featured, in addition to a
fascinating, new (to me) argument for the value of the famous number
$33/{12\pi}$ of QCD, a bold new idea on the color structure of the
population of final-state soft gluons in the lego plot, namely that the
color-lines connecting the emitted gluons wrap themselves tightly around
the (cylindrical) lego phase space. It is so much fun that I cannot help
but return to it in the next section, and attempt to describe it in my own
language.

{\em Gluon jets} (Gary\cite{**}): The DELPHI analysis of gluon jets tagged in
$b-\bar {b}$ events continues to provide a supremely pure gluon-jet
subset.  They provide a superb standard of reference for basic
gluon-fragmentation properties.

{\em Rapidity gap events}: Here there is steady progress from HERA and the
Fermilab Tevatron, which I will not attempt to summarize. To me the most
important news is that in the most recent D0 data on gaps between jets,
the evidence seems to favor the color-evaporation mechanism described here
by Halzen\cite{**}, and not the perturbative two-gluon-exchange mechanism
(plus spectator effects) advocated by myself and others. If this holds up,
there will be no known hard diffraction process where perturbative-QCD
mechanisms are adequate for the description, however large the momentum
transfer or virtuality of the projectiles. Again, I will return with an
additional comment on this subject in the next section.

{\em Observation of an exotic meson} (Seth\cite{**}; BNL E852): The evidence for
a broad $1^{-+}$ state with mass about 1350 MeV, decaying into $\eta\pi$,
looks quite good.  Its quark-gluon structure should be either $q\bar{q}g$
or $q\bar{q}q\bar{q}$ and is the first of its kind to be established. It
again is a strong reminder of how much remains to be done in the field of
light-quark spectroscopy, and how much more might be accomplished with a
new, totally modernized, dedicated effort. 

{\em Regge theory for heavy flavors} (Srivastava\cite{**}): The question of how
Regge trajectories for mesons which are built from at least one heavy
quark are to be extrapolated into the spacelike regime is a very nice
problem, and is very nicely addressed in this contribution. It was all new
to me.

{\em FELIX} (Eggert\cite{**}): This initiative, to provide the LHC with a
full-acceptance detector dedicated to investigating all aspects of QCD and
multiparticle dynamics in hadron-hadron collisions, is of great potential
importance to this field. I return to it at length in Section 4.

\section{RANDOM ITEMS}

This section consists of comments on the following topics: (1) the pushing
of perturbative-QCD techniques to very low momentum scales, (2) Bo
Andersson's screwy gluons, (3) the nature of inelastic diffraction, and
(4) wee partons in onium-onium scattering.

\smallskip\noindent
1. {\em Perturbative QCD in the infrared}: 

 It has been a popular and rather successful activity for some time to use
short-distance techniques for the study of multiparticle production.
Parton cascades are carried to very low momentum scales, below 1 GeV, with
the distributions of partons then identified with the distributions of
produced hadrons. This is very close to what is done in the Monte Carlo
simulations, for practical reasons, and now we see analytic calculations
emulating the simulations. Yuri Dokshitzer provides in these
proceedings\cite{**} a review and something of a critique of this
approach.

I am becoming more and more uncomfortable with all this\cite{bj1}. It is not that
this kind of thing should not be done, but that there needs to be a very
cautious and critical attitude regarding what it means. It is not enough
to fit the data; the significance of the agreement needs to be critically
addressed. 

Important tools in this approach are the notions of infrared safety, local
parton-hadron duality, and preconfinement. Infrared safety means choosing
variables which are perturbative-QCD user-friendly. Local parton-hadron
duality means, in this context, that the partons produced in the QCD
cascade are essentially in one-to-one correspondence with produced
hadrons. Preconfinement means that the color pattern of produced partons
is such that in the final stage, after gluons are (by hand) replaced by
colored $q \bar{q}$ pairs, the quarks and antiquarks can be recombined
into hadrons locally in phase-space almost all the time, thanks to the
color structure of the QCD cascade. 

Infrared-safe is not the same as infrared-accurate. An easy example is
that exemplar of perturbative QCD, the infrared-safe $e^{+}e^{-}$
cross-section ratio $R$.  The perturbative $R$ for a given flavor
increases monotonically with energy while the data rides up and down over
resonances. Locally there is little chance of getting an accurate value
for $R$ until $s$ is above 2-4 GeV$^2$ for light flavors. Global averaging
is needed over a GeV scale to retrieve safe predictions. And I think it
especially important to recognize that the criterion for accuracy is {\em not}
to be obtained from perturbation-theory estimates, but requires knowing 
something about the nonperturbative sector, i.e. the resonances. 

In the case of $R$ and its close relatives, one can do more, such as the
QCD sum-rules. But these methods lean heavily upon the simple analyticity
properties of the correlators, something not present for most applications
in the multiparticle-production world. 

Thanks in good part to the preconfinement feature of QCD cascading,
pushing the effective hadronization scale to very low values does work on
average rather well for gross features of multiparticle production. But
again we may look at $e^{+}e^{-}$ annihilation to question the
significance.  The average properties of the final multihadron final state
can indeed be accounted for at low invariant masses economically using the
perturbative-QCD cascade and local duality. But below $s$ of 20--30 GeV$^2$
there are no discernible jets in to be seen in the final state. Isotropic
phase space also does quite well in accounting for the data. In the
1970's, before the ``right answer" was found, there were a plethora of
competing interpretations of such data. If a QCD cascade is used just to
generate a phase-space distribution, there is no reason to hail it as
another triumph of QCD. However, if the same cascade is used in a Monte
Carlo simulation, it may be hailed as a triumph of efficient use of
computer time. 

More interesting to me is the question of limitations of the method: they
do exist. In particular the perturbative cascades \textit{predict}
occasional non-confinement. For example, even at the $Z$ mass there is a
small but nonvanishing probability that no gluons are emitted during the
cascade and that one is left at the end of the line with only the original
$q\bar{q}$ pair. (The probability for this, if local duality works, might
be guessed to be of order the branching ratio for $Z\rightarrow\pi\pi$.)
But at the hadronization time, when the QCD cascading is terminated, the
$q$ and $\bar{q}$ are very far apart, and there is nothing to be done
about that issue using perturbative means alone. To me the message here is
that the low-multiplicity tails of the sundry distributions are likely to
be the most sensitive to deficiencies of the perturbative-cascade
technique, and therefore may be the most important to study incisively.
The bunch-parameter method described here by Chekanov\cite{**} looks 
like a good way to approach the problem. 

\smallskip\noindent
2.  {\em Screwy gluons}: 

Here is a paraphrase of Bo Andersson's talk\cite{**}. It is what I would
have him say, but probably not what he would have it be. The deficiencies
here are my responsibility, of course. 

At the end of a typical perturbative-QCD cascade, now in the higher
multiplicity regime, the lego plot will be rather crowded with produced
gluons, most all of which will have $p_t$ of order the infrared cutoff.
(Neglect quarks, other than  leading quarks, and consider again for simplicity the
$e^{+}e^{-}$ process.)  Consider also the large-$N_c$ limit; then for a
given event there is one color line which flows through the gluon
population linking them together. The amplitude for the production process
is largest when gluons connected by the color line have low subenergies,
i.e. are close together in the lego plot. Just from massless kinematics,
this (squared)  subenergy contains a factor ($\cosh \Delta\eta - \cos
\Delta\phi$). For small separations, this is isotropic, just the square of
the (lego)  distance between the gluons. (I consider a single gluon to
occupy a region of phase space which is a circle of radius 0.7; if the gluons 
are separated by much less than 1.4 lego units, they should be considered as 
merged into a single collinear gluon of approximately twice the $p_t$.)  
Now comes the main point: the aforementioned isotropy is only approximate, 
and for a fixed distance between gluons, the subenergy is minimized when 
the two gluons have the same $\eta$ and different $\phi$, rather having different 
$\eta$ and the same $\phi$. If this is demanded for all gluon pairs connected by a color
line, then one sees that this must lead to the color line spiraling around
the (cylindrical; $\phi$ is periodic!) lego plot.

I worry that the amplitude for producing this configuration, while
arguably maximal in size, is a very special one, and that all the other
somewhat similar amplitudes, where the color line wanders rather randomly amongst
nearest-neighbor gluon pairs, will overwhelm this special one. In other
words the color-line entropy in this Feynman-Wilson gluon fluid may be
more important than its energy. 

However, the bottom line, independent of the details, is that there has
been identified a new \textit{winding number} characterizing multiparton
``final states". Whether it is maximal or has a broad distribution is less
important than its existence. Of course it is essential to ask the
question of what signatures in the final-state hadron distributions are
implied. I am sure this question is being asked in Lund, and I very much
look forward to the answer.

\smallskip\noindent
3. {\em Soft diffraction}: 

Nowadays in the field of single diffraction, especially in the relatively
new subfield of hard diffraction, there is used the formalism of the
exchange of a Pomeron Regge pole. I am very uncomfortable with this
situation, if only that the basic observational data are often presented in
this very model-dependent fashion. Furthermore, as emphasized especially
by Goulianos\cite{**}, the phenomenology of soft single diffraction does
not support Regge-pole exchange, because the ratio of single diffraction
dissociation to a given fixed mass interval to elastic scattering should
be constant with increasing energy in the Regge-pole picture, while
experimentally it clearly decreases.

While there is still room for debate on this point, it remains that there
is at present great uncertainty in understanding what is going on in soft
diffraction dissociation to high mass systems. There is every reason to be
even more open-minded on what is happening in the many hard diffraction
processes, which all seem so far to involve soft-Pomeron and/or soft QCD
mechanisms in one way or another. 

My own choice for meeting point between theory and experiment is the
determination of the {\em gap fraction}, i.e. the probability per
inelastic event, at the same cms energy and same gross event properties
(such as $Q^2$ for electroproduction and jet locations and $p_t$'s for
hadron collisions), of finding a rapidity gap with gap edge in some
prescribed location. (At HERA this is about 4--5\%\  per unit rapidity,
while at hadron colliders it is 0.5--1\%.) I am still working on this
proposition, and am sorry it is not yet ready in detail to present here. 

Thinking in terms of gap fractions evokes a somewhat different picture of
the diffraction process, more in line with the original picture of
diffraction dissociation created by its inventors, Good and
Walker\cite{GW}. When two hadrons pass through each other in a peripheral
collision, the internal wave functions will be modified because there is
more absorption on the overlapping parts of the disks than on the opposite,
nonoverlapping parts. So just after the disks pass through each other, the
internal wave functions will neither be the ground state hadron wave
functions nor be orthogonal to them. At a later time, God throws the dice
and either the ground states are chosen (elastic scattering), one ground
state and one excited state is chosen (single diffraction), or only
excited states are chosen (double dissociation or a fully inelastic
interaction). The main point is that in this picture, the ``decision"
appears in the future--more precisely within the future light cone for the
collision process. The diffraction process is not ``prepared", before the
collision occurs, by the nucleon ``emitting a virtual Pomeron, which then
collides with ...".  This is a nontrivial distinction, since a photon- or
gluon-exchange picture of the process does imply, a la the
Weizsacker-Williams viewpoint, just such a ``preparation". To be sure, even
in the Good-Walker picture the probability that diffraction occurs will be influenced by 
initial-state quantities such as impact parameter, but this does not seem to me to be
the same thing as the Weizsacker-Williams picture. 

In any case the global space-time picture of single diffraction is to me a
fascinating one. For example, in what we have described above, the
decision in the future is that what starts out as a candidate ``elastic" 
process ends up, thanks to God's roll of the dice, as high-mass
diffraction. The ``decision" is made on a large time scale, proportional to
the energy of the softest particle produced in the diffracted system. On
the other hand, view this same event in a boosted reference frame,
in particular the center-of-mass frame of the produced massive diffractive
state. In that frame the process starts out as an ordinary {\em inelastic}
collision, but where at a certain time in the future (proportional to the
fastest left-moving particle within the high-mass diffractive system),
God's roll of the dice \textit{stops} the left-moving particle production
and leaves only a single unexcited projectile on the far side of the
rapidity gap.  So the physics of creation of a rapidity gap in the course
of a collision must be essentially the same as that of the termination of
one already present. And this is something which happens on very long time
scales.

How this works in detail is clearly subtle, no matter what one's opinion
of the nature of high-mass soft diffraction. I think there is still a
great deal to understand about this process. 

\smallskip\noindent
4. {\em Onium-onium scattering and the limitations of the BFKL picture}:

The relevance of the BFKL formalism of high-energy scattering of colored
partons, in particular of small color dipoles, to actual scattering
phenomenology is still controversial and uncertain. The critical view was
expressed rather strongly in Bo Andersson's splendid and provocative
summary of last year's conference\cite{bo}. I too feel uneasy about its
relevance.  I have also been for a very long time uneasy about
the Fock-space picture of hadron wave functions used extensively for the
description of exclusive and semi-exclusive high energy processes. There
are many who argue that in an infinite-momentum proton there is a finite
probability for finding a finite number of partons, because if the valence system forms a
small-size configuration there seems to be no appreciable  gluon field
available to create the nonperturbative
component. On the other hand the mean number of partons is measured to be
infinite, unless the deep inelastic $F_2$ eventually goes to zero at very
small $x$, something nobody expects to occur. In this Fock-space viewpoint,
the multiplicity distribution of partons in a very energetic nucleon will
not completely recede to large values as the energy tends to infinity, but will 
have a residual piece at finite multiplicity. What fraction of the distribution is 
in this piece? What is the probability of finding a finite number of partons in an
infinite-momentum projectile? 

To me there is a clear answer to the above questions, namely that there is
always an infinite number of the wee partons. Nevertheless there is still
a great deal that is right about the Fock-space way of thinking and
therefore a problem to address. Recently I have come across a point of
view which at least to me is helpful, and that is the reason for these
paragraphs.

First of all, the origin of the usual primordial distribution of
small-$x$, low-$Q^2$ partons is probably not to be found in perturbative
mechanisms. There have been many such attempts, but to the best of my
knowledge they just don't work. Small-$x$, low-$Q^2$ partons arguably
belong to the physics of the soft Pomeron, i.e. the same physics
responsible for the high energy behavior of the total $pp$ cross section.
This, in turn, is associated with large impact parameter collisions, i.e.
the collisions of pion clouds with each other. So I assume that at least
some part, if not all, of the primordial wee-parton distribution is to be
associated with the internal structure of the pion clouds surrounding
hadrons.

Now consider the BFKL process of the scattering of two small color
dipoles, such as virtual photons. They are splendid candidates for a
Fock-space configuration; I am not worrying here (yet) about the
corrections to this configuration, calculable in principle from
perturbative QCD, which might give a very dilute BFKL-like distribution
at small $x$. Instead I concentrate on the presence of a very weak pion
cloud around the color dipole. In an onium rest frame (or something like
it for virtual photons), the small color dipole must introduce a small
distortion in the structure of the neighboring vacuum, i.e. chiral
condensate. This is another way of saying there must be a pion cloud
around even these perturbative structures, whose strength is measured by
the strength of the dipole. There will be an energy associated with this
distortion. Even if it is minuscule in the onium rest frame, it becomes a
large momentum density when the onium is boosted to high enough energies.
And when this momentum density of the Lorentz-contracted disk exceeds, say
1 GeV/fm$^2$, the projectile will interact \textit{strongly} with
stationary hadrons or other onia moving in the opposite direction with
momentum density also exceeding 1 GeV/fm$^2$. At this point, the
perturbative description will not work. Onium-onium scattering will fail
to be described by a BFKL picture at all. 

The parton distributions for the small onia will therefore be of the
Fock-space character at large $x$, while at sufficiently small $x$ there
will be a transition to the parton distributions of the boosted pion
cloud. The critical $x$ at which this occurs will be proportional to the
square of the dipole moment of the onium or virtual photon. In particular,
the critical $x$ in deep-inelastic scattering below which this occurs will
be inversely proportional to $Q^2$.

This is a rather strong conclusion, so strong that I find myself
surprised, and therefore still cautious. If one accepts this point of
view, it would seem that the Gribov bound in electroproduction\cite{bj2}
would be satisfied for a rather large domain in $x$ and $Q^2$. 

I was questioned by Bartels as to whether this picture is essentially just
the same as the well-known phenomenon of the $p_t$ diffusion of the rungs
of the BFKL ladder in transverse-momentum space. This phenomenon also
leads to the eventual breakdown of the BFKL picture, and the inevitable
mixing of the perturbative Pomeron with the soft Pomeron. I think there is
a distinction to be made, since the diffusion argument is made from the
short-distance side of QCD, while the pion-cloud argument starts from the
large-distance limit of QCD.  The difference is similar to approaching
intermediate-energy $e^{+}e^{-}$ annihilation physics from the
perturbative QCD or QCD sum-rule approach, rather than looking at it from
the point of view of resonance spectroscopy and chiral perturbation
theory. Both approaches may identify the same kinematic boundary, but they
are certainly not to be regarded as identical. 

\section{FELIX: A QCD DETECTOR FOR THE LHC}

Karsten Eggert\cite{**} presented here a description of a new initiative,
FELIX, for the study of QCD and multiparticle production at the LHC. The
FELIX detector, with full acceptance, would be dedicated to the
comprehensive study of both hard and soft QCD. At present there are the
beginnings of an experimental collaboration, and a Letter of Intent (LOI)
has been produced and submitted to the CERN LHC Committee\cite{LOI}. This
LOI is a rather long letter, consisting of about 200 pages of details, of
which more than 70 have to do with the theoretical justification. A very
large number of QCD theorists have in fact contributed to the document,
and what emerges is a remarkable view of a quite different phenomenology
emergent at the LHC energy scale, as well as a rich variety of fresh ideas
on what novel QCD processes would be of interest to measure if only there
were available a good enough instrument for doing so at hadron collider
energies.

It seems to me that there is no better way to express an optimistic
outlook for the future of this subfield than to directly pirate some of
the material from this LOI. Without further ado, the question to be
answered is why one should build FELIX to study QCD: 

\smallskip\noindent
1. {\em QCD is universal}: 

There is precious little in particle physics that does not depend upon
QCD. Take away quarks, gluons and other conjectured colored particles and
there is not much left. And while one may not be motivated by wanting to
understand QCD better if one is intent on discovering the gluino, it nevertheless
clearly helps to do so in making such a search. 

\smallskip\noindent
2. {\em Dedicated study of QCD at the LHC is essential}: 

This community of specialists in multiparticle dynamics is most aware of
the deficiencies of the QCD database for hadron-hadron collider processes. 
It also is most aware of the limitations of theory and of simulation tools in
adequately anticipating what will occur in even the most generic of
collisions, not to mention the properties of the more unusual classes of
events. This richness of the phenomenology, together with the importance
of understanding it, if for no other reason than to optimize the analysis
of new-physics processes, demands instrumentation, and the dedicated human
effort behind the instrumentation, which will do an optimal job of
understanding the QCD mechanisms underlying the properties of all final
states to be observed at the LHC.

\smallskip\noindent
3. {\em Minijet production is strongly energy dependent}: 

As discussed here by Giovannini\cite{**} and by Treleani\cite{**}, parton
densities at the LHC energy scale become so high that minijet production
in generic central collisions becomes commonplace. Almost all such central
events are expected to contain observable minijets in the 10 GeV range of
$p_t$. These very high parton densities create at a short distance
perturbative-QCD scale, ``hot spots" of very large local energy densities.
There is the real possibility that these evolve in nonperturbative ways,
creating local thermalization and/or other phenomena not easy to
anticipate in advance. Particle spectra may be enriched in heavy flavors
and/or baryons and antibaryons.  Especially in central proton-ion
collisions--accessible at the LHC and observable with the FELIX
detector--the gluon-gluon luminosity will be maximized. The evolution of
the proton fragments can be followed, and one can anticipate that this
class of phenomena may be especially rich in surprises.

\smallskip\noindent
4. {\em Parton densities can be measured to extremely low $x$, below $10^{-6}$}: 

FELIX can measure parton densities at very low $x$ via the production of
leading dilepton, dijet, and $\gamma$-jet systems with masses in the 5-20
GeV range, and laboratory momenta in the 1-5 TeV range. In this regime one
expects, especially in proton-ion collisions, the breakdown of the usual
BFKL/DGLAP evolution equations, and significant nonlinear effects to
occur. 

\smallskip\noindent
5. {\em Diffractive final states are endemic, many are important, and some are
   spectacular}: 

The contributions to this meeting attest to the fact that the rather young
field of hard diffraction is one of great interest and vitality, limited
mostly by the present energy scale and by the absence of detector
acceptance in the far forward and backward regions. The interest in hard
diffraction in turn has focused more attention on soft diffraction as
well, which as discussed above is not well understood either.

It is not acceptable scientifically that a set of processes which at the
LHC will comprise almost half of all interactions, and which occurs even
in rare multijet events at the one percent level, remain not understood.
And given the expectation that for example, at an eminently measurable
rate, there will exist events containing a central high-$p_t$ dijet, two
Roman-pot protons carrying the beam energies, and absolutely nothing else
in the detector, there can be no excuse for not pursuing their observation
and interpretation with the utmost of vigor. 

The major detectors at the LHC can and should observe such final states,
provided they sacrifice their hard-won luminosity by a factor 30 or so.
However to really understand such events one will want to determine the
$t$-dependence of Roman-pot protons, as well as to study the
generalizations of the process to the large-$t$ regime where one or both
beam protons fragment. Only FELIX would have the capability for such
studies. 

There are a myriad of basic diffractive processes to study, as was
exhibited in Eggert's talk, and FELIX will be a superb instrument for
doing so. In addition, the $A$-dependence of diffraction is of great
usefulness as a diagnostic of the underlying mechanisms, and that will be
available as well.

In addition to the class of hard and soft diffraction into high-mass
systems, there is another very interesting class of semihard processes
associated with the conjectured fluctuation of the projectile into a
transversely compact configuration, which therefore interacts with a small
cross section. Electroproduction of vector mesons, especially $\psi$'s,
shows the rapid increase with energy expected. Coherent dissociation of a
pion into a dijet pair from nuclei, apparently seen in the E791 experiment
at Fermilab\cite{791}, is another example, one which can be reproduced at
FELIX--even including the nuclear target. Another variant is coherent
dissociation of a proton projectile into trijets. 

\smallskip\noindent
6. {\em Particle production from deep inside the light cone may exist and
deserves careful searches}: 

Almost all generic particle production models, whether parton-cascade,
string-fragmentation, or multiperipheral, put all the nontrivial activity
in spacetime near the lightcone. However, some of us have speculated that
the interior of the future lightcone might contain disoriented vacuum, or
chiral condensate (DCC), the pionic decay products of which have large
Centauro-like fluctuations of the neutral-to-charged ratio. So far
searches for this, carried out by the MiniMax group at Fermilab and by
the WA98 group at CERN, reported here by Peter Steinberg\cite{**}, have
not turned up anything out of the ordinary. But this is a young subject
and the techniques are still primitive, both theoretically and
experimentally. 

There is plenty of room for skepticism regarding whether the DCC picture
is realistic. In particular one can question why the interior of the light
cone should relax to vacuum rapidly at all. If it does not, then there
must be new mechanisms of particle production associated with its decay at
large proper times. And if it does promptly relax to vacuum, I see no
strong argument why it chooses the chiral order parameter of the exterior
vacuum, given the hot shell in between and the short time scale for
sensing the small chiral symmetry breaking. 

The bottom line is that the searches for novel phenomena along these lines
are worth doing. The technique at the least has to include isolation and
precise measurement of low $p_t$ photons and charged secondaries,
preferably with particle identification, in high associated-multiplicity
environments (including minijet interiors). All but the
particle-identification requirement is met by FELIX. 

\smallskip\noindent
7. {\em Collisions with large impact parameter may probe the chiral structure
of QCD}: 

Large impact-parameter collisions by definition involve the passage of the
pion clouds of the projectiles through each other, suggesting that a
connection with LQET might eventually be established. Is the physics of
this ``supersoft Pomeron" linked somehow with the chiral sector of QCD?
Study of this class of final states, especially nondiffractive, has not
been done, especially given the lack of an experimental impact parameter
tag. But as discussed below, FELIX offers the best opportunity in a long
time for producing one. Emphasis on low mean multiplicity, low $p_t$, and
leading-particle properties is a requisite in the exploration of this
event class.

\smallskip\noindent
8. {\em New opportunities exist for tagging event classes}:

Determination of impact parameters event-by-event, as desired above, is
routinely attained in heavy ion collisions, and is an essential analysis
tool. It is determined by observation of the forward fragments, and by
determination of the transverse energy at central rapidities. In
hadron-hadron collisions at collider energies, such techniques are
essentially nonexistent at present. There have been no forward detectors,
and the fluctuations in transverse energy event-by-event due to other
mechanisms have been too large. However with the prevalence of minijet
production at the LHC energy scale, there is, as discussed here by
Treleani\cite{**}, the opportunity of using the latter method, while FELIX
would provide the former. The actual algorithms will have to be determined
by a data driven approach, and of course success is not guaranteed.

There are many other categorizations of event classes in $pp$ collisions. An
especially interesting set is to convert the incident proton beam into a
pion beam by isolation of the one-pion-exchange contribution via tagging
or triggering on a leading neutron or $\Delta^{++}$. Kaon beams can be
made via a leading $\Lambda$ tag, and of course ``Pomeron" beams with Roman
pot tags of leading protons. In addition an important portion of the FELIX
program is to tag on Weizsacker-Williams photons (via the nonobservation
of completely undissociated forward ions) in ion-ion running, creating a
high-luminosity $\gamma-\gamma$ collider. 

In addition there are pattern tags. The acceptance of FELIX is so large
that the texture of the final state hadronic system is itself a signature. 
Beyond the very strong textures of jets, rapidity gaps, and mean particle
densities, there may be observable the effects of color-flow (``antenna
patterns") and/or fluctuation structures in the lego plot associated with
intermittency, DCC production, etc. Working at this level will be
challenging, requiring close interplay between the experiments and the
phenomenology. But the extremely large lego acceptance of FELIX and the
richer underlying physics at the LHC energy scale should make these
opportunities much stronger. 

\smallskip\noindent
9. {\em Benchmark processes}:

Because the FELIX program is so broad, it has been decided to focus on a
few ``benchmark" processes for detailed analysis. Candidates for these have
been chosen, with an eye to exhibiting the variety of physics goals as
well as to exercise the detector capability as broadly as possible. These
include the following: 

\begin{itemize}
\item
[a)] {\em Elastic scattering at large $t$}: 

This has not been done since the ISR and clearly exercises the Roman pot
capability. 

\item
[b)] {\em Spectra of leading particles}: 

This is straightforward physics, of importance especially to the cosmic
ray community, and exercises the FELIX forward spectrometer. 

\item
[c)]  {\em Mueller-Navelet jets}: 

Dijets of laboratory energy of, say, 3 TeV will be measured from the
lowest $p_t$ scale of 5-10 GeV out to large angles, along with the event
structure (extra jets, rapidity gaps, etc.). At small angles the jet
reconstruction will be an experimental challenge.

\item
[d)] {\em Parton distributions at small $x$}: 

This is the aforementioned measurement of low mass dileptons, dijets, and
jet-photon systems at large laboratory energies.

\item
[e)] {\em Hard double Pomeron exchange}: 

This process is characterized by four jets and two or three rapidity gaps,
and exercises the capability of pattern triggers.

\item
[f)] {\em Tagged pion dissociation into dijets}: 

We have already mentioned the physics interest. Experimentally this is
very challenging, because it may require putting Roman-pot signals into
the trigger, and because beam-gas backgrounds may be troublesome. 

\item
[g)] {\em Proton diffraction into three jets from nuclei}: 

The physics is similar, but here the problems involve how the products of
a dissociated ion are seen in the FELIX detector. It is probably even more
difficult than the previous case. 

\item
[h)] {\em Vector meson production in gamma-gamma collisions}: 

There is a serious problem of backgrounds from Pomeron-exchange processes. 
\end{itemize}

It seems very clear to me that FELIX presents an extremely important
opportunity for advancing the field of multiparticle dynamics. I hope that
everyone will familiarize themselves with the enterprise, will spread the
word, and if possible will contribute to the effort in one way or another.
The best of course would be to join the collaboration. Short of that,
there are many ways of helping the project along, both on the theoretical
side and experimental. I urge you to contact the spokespersons, Karsten
Eggert and Cyrus Taylor, and/or visit the FELIX web site
(http://www.cern.ch/FELIX) and join the working group. It will be at best
a very hard job to make FELIX a reality, and cohesive support from a broad
sector of the community will be essential.

\bigskip
\noindent
{\bf NOTE ADDED IN PROOF:}
\bigskip

On November 5--6, the CERN LHCC acted in a nonpositive manner regarding the FELIX 
LOI\cite{nine}.  I find especially noteworthy their statement that ``while the physics topics
addressed by the programme proposed in the LOI are of interest (particularly the complete 
reconstruction of diffractive events), the likely costs of constructing the proposed dedicated 
detector and of the modifications to the LHC collider are very high in comparison with the
probable physics output.''  I strongly disagree with this assessment.

\section*{ACKNOWLEDGEMENT}

On behalf of all the participants, I thank Giulia Pancheri and all the
organizers for their hard work in creating a most pleasant and productive
meeting.


\begin{thebibliography}{9}

\bibitem{bj1} 
Very similar comments can be found in my summary talk from the 1994 conference:
J. Bjorken, {\sl XXIV International Symposium on Multiparticle Dynamics}, 
Vietri sul Mare, Italy, Sept. 1994, ed. A. Giovannini, S. Lupia, and R. Uguccioni 
(World Scientific), p. 579.

\bibitem{ss}
T. Schaefer and E. Shuryak, hep-ph/9610451; {\sl Rev. Mod. Phys.} (to be published).

\bibitem{**} 
Contribution to these Proceedings. If you are a contributor
and are looking for your name, you are invited to peruse the full text of
this talk to find it.

\bibitem{GW} M. Good and W. Walker,  {\sl Phys. Rev.} {\bf 120}, 1857 (1960).

\bibitem{bo} 
B. Andersson, {\sl XXVI International Symposium on Multiparticle Dynamics}, 
Faro, Portugal, Sept. 1996, ed. J. Dias de Deus, P. Sa, M. Pimenta, S. Ramos, 
and J. Seixas (World Scientific), p. 471.

\bibitem{bj2} 
See for example, reference 1 for a discussion.   Note that this point of view is more 
or less expressed by Fig. 4(a).  In that talk I favored Fig. 4(b).  At least something has 
changed in my summary talks.

\bibitem{LOI} 
FELIX : A full acceptance detector at the LHC (Letter of Intent); CERN/LHCC 97-45; 
LHCC/110; August, 1997.

\bibitem{791} 
R. Weiss-Babai; presented at the Hadron'97 Conference, Brookhaven National Laboratory. 

\bibitem{nine}
The full text can be found in \hfill\break http://www.cern.ch/Committees/LHCC/
\hfill\break LHCC31.html.

\end{thebibliography}
\end{document}